\documentclass{ws-p10x7}
\usepackage{epsfig}
\usepackage{epsf,psfrag}

\newcommand{\MD}{\ensuremath{\mathcal{D}} }
\newcommand{\MK}{\ensuremath{\mathcal{K}} }
\newcommand{\MU}{\ensuremath{\mathcal{U}} }

\newcommand{\MC}{\ensuremath{\mathcal{C}} }

\newcommand{\omD}{\omega_{\MD}}

\begin{document}

\title{Infrared freezing of Euclidean QCD observables in the one-chain approximation}

\author{C.J. Maxwell}

\address{Institute for Particle Physics Phenomenology, Durham University, Durham, DH1 3LE, UK
\\E-mail: c.j.maxwell@durham.ac.uk}

\twocolumn[\maketitle\abstract{We consider the one-chain term in a skeleton expansion for Euclidean
QCD observables. Focusing on the particular example of the Adler $D$ function, we show that
although there is a Landau pole in the coupling at $Q^2={\Lambda}^{2}$ which renders
fixed-order perturbative results infinite, the Landau pole is absent in the all-orders
one-chain result. In this approximation one has finiteness and continuity at $Q^2={\Lambda}^{2}$, and a 
smooth freezing as $Q^2\rightarrow{0}$.  }]

\noindent In this talk I want to describe some recent work with Paul Brooks \cite{r1} in
which we consider the low-energy behaviour of Euclidean QCD observables. We investigate
the $Q^2$-dependence of all-orders perturbative resummations obtained using the so-called
leading-$b$ approximation \cite{r2,r3,r4}, where $b=(33-2N_f)/6$ is the leading beta-function coefficient
in SU($3$) QCD with $N_f$ active quark flavours. This is closely related to the one-chain term
in a skeleton expansion in which a single chain of fermion bubbles is inserted into a basic skeleton
diagram \cite{r5,r9}. We shall demonstrate that in this approximation the Landau pole in the coupling at
$Q^2={\Lambda}^{2}$ is absent, and one has finiteness and continuity at this energy, with a smooth 
freezing behaviour at lower energies. We focus in this talk on the Adler $D$-function, but DIS sum
rules are also studied in Ref.\cite{r1}.\\

\noindent The Adler $D$ function is directly related to the vacuum polarization function ${\Pi}^{\mu\nu}(Q^2)$,
($Q^2=-q^2>0$),
\begin{equation}
{\Pi}^{\mu\nu}(Q^2)=16{\pi}^{2}i\int{d^4x}{e}^{iq.x}\langle{0}|T({J}_{\mu}(x){J}_{\nu}(0))|0\rangle.
\end{equation}
Current conservation dictates that this has the tensor structure
\begin{equation}
{\Pi}^{\mu\nu}(Q^2)=({q}_{\mu}{q}_{\nu}-{g}_{\mu\nu}{q}^{2}){\Pi}(Q^2).
\end{equation}
Only $\Pi(Q^2)-\Pi(0)$ is observable, so it is useful to eliminate the constant and define
the {\it Adler Function} $D(Q^2)$
\begin{equation}
D(Q^2)=-\frac{3}{4}Q^2\frac{d}{dQ^2}{\Pi}(Q^2).
\end{equation}
This may be written as a sum of the parton model result and QCD corrections ${\cal{D}}(Q^2)$
\begin{equation}
D(Q^2)=3\sum_{f}{Q}_{f}^{2}[1+{\cal{D}}(Q^2)].
\end{equation}
These QCD corrections are split into a perturbative and non-perturbative part
\begin{equation}
{\cal{D}}(Q^2)={\cal{D}}_{PT}(Q^2)+{\cal{D}}_{NP}(Q^2).
\end{equation}
The PT component has the form
\begin{equation}
{\MD}_{PT}({Q}^{2})={a}({Q}^{2})+{\sum_{n>0}}{d}_{n}{a}^{n+1}({Q}^{2}).
\end{equation}
Throughout the talk we will take $a(Q^2)\equiv{\alpha}_{s}(Q^2)/\pi$ to be the one-loop form of the 
QCD coupling
\begin{equation}
a(Q^{2})=\frac{2}{b\ln(Q^{2}/\Lambda^{2})}\;.
\end{equation}
The NP component in Eq.(5) will have the form
\begin{equation}
{\MD}_{NP}({Q}^{2})={\sum_{n}}{\MC}_{n}{\left(\frac{{\Lambda}^{2}}{Q^2}\right)}^{n}\;.
\end{equation}
The leading OPE contribution for the Adler function is the dimension 4 gluon condensate contribution
\begin{eqnarray}
{G}_{0}(a(Q^2))=\frac{1}{Q^4}\langle 0|GG|0\rangle{C}_{GG}(a(Q^2))\;.  
\end{eqnarray}
We are interested in the behaviour of ${\cal{D}}(Q^2)={\cal{D}}_{PT}(Q^2)+{\cal{D}}_{NP}(Q^2)$ as $Q^2\rightarrow{0}$.
Clearly at any fixed order perturbation theory breaks down at $Q^2=\Lambda^2$, the Landau pole in the coupling, and 
$a(Q^2)\rightarrow{\infty}$. Evidently we need a resummation of perturbation theory to all-orders to address the freezing
question, and we need to combine the resummation with the OPE condensates. The large-$N_f$ limit provides a way of
formulating this resummation.
The coefficient $d_n$ may be expanded in powers of $N_f$ the number of quark flavours
\begin{equation}
{d}_{n}={d}_{n}^{[n]}{N}_{f}^{{n}}+{d}_{n}^{[n-1]}{N}_{f}^{n-1}+\ldots+{d}_{n}^{[0]}
\end{equation}
The leading large-$N_f$ coefficient ${d}_{n}^{[n]}$ may be evaluated to all-orders
since it derives from a 
restricted set of diagrams obtained by inserting a chain of fermion bubbles 
inside the quark loop. The crucial ingredient in the calculation is the chain
of $n$ renormalised bubbles ${B}^{\mu\nu}_{(n)}$,
\begin{eqnarray}
{B}_{(n)}^{\mu\nu}=\frac{(k^2{g}^{\mu\nu}-{k}_{\mu}{k}_{\nu})}{{(k^2)}^{2}}{\left[-\frac{N_f}{3}\left(\ln\frac{k^2}{\mu^2}+C\right)\right]}^{n}.
\end{eqnarray}
Here $k$ is the momentum flowing through the chain of fermion bubbles.
The factor in the square bracket is the one-loop vacuum polarization contribution ${\Pi}_{0}(k^2)$.
The constant $C$ depends on the subtraction procedure used to renormalise the bubble. With
$\overline{MS}$ subtraction $C=-\frac{5}{3}$. We shall choose to work in the ``V-scheme'' which corresponds
to $\overline{MS}$ with the scale choice $\mu^2={e}^{-5/3}Q^2$, in which case $C=0$. The result for ${d}_{n}^{[n]}(V)$ is \cite{r5,r6}
\begin{eqnarray}
&{d}_{n}^{[n]}(V)&=\frac{-2}{3}(n+1){\left(\frac{-1}{6}\right)}^{n}\left[-2n-\frac{n+6}{{2}^{n+2}}\right.
  \nonumber \\
&+&\frac{16}{n+1}{\sum_{\frac{n}{2}+1>m>0}}m(1-{2}^{-2m})
\nonumber \\
&\times&\left. (1-{2}^{2m-n-2}){\zeta}_{2m+1}\right]{n}!\;.
\end{eqnarray}
This large-$N_f$ result can describe QED vacuum polarization, but for QCD the corrections
to the gluon propagator involve gluon and ghost loops, and are gauge-dependent. The result
for ${\Pi}_{0}(k^2)$ is proportional to $-N_f/3$ which is the first QED beta-function coefficient, $b$. 
In QCD one expects large-order behaviour of the form ${d}_{n}\sim K{n}^{\gamma}{(b/2)}^{n}n!$ involving
the QCD beta-function coefficient $b=(33-2N_f)/6$, it is then natural to replace $N_f$ by $(33/2-3b)$ to
obtain an expansion in powers of $b$ \cite{r2,r3,r4}
\begin{eqnarray}
{d}_{n}={d}_{n}^{(n)}{b}^{n}+{d}_{n}^{(n-1)}{b}^{n-1}+\ldots+{d}_{n}^{(0)}
\end{eqnarray}
The leading-$b$ term ${d}_{n}^{(L)}\equiv{d}_{n}^{(n)}b^n={(-3)}^{n}{d}_{n}^{[n]}b^n$ can
then be used to approximate $d_n$ to all-orders, and an all-orders resummation of these terms
performed to obtain ${\cal{D}}^{(L)}_{PT}(Q^2)$. 
If we use the Borel method to define the all-orders perturbative result we obtain
\begin{eqnarray}
{\cal{D}}^{(L)}_{PT}({Q}^{2})={\int_{0}^{\infty}}{dz}\,{e}^{-z/a({Q}^{2})}B[{\cal{D}}^{(L)}_{PT}](z)\;.
\end{eqnarray}
The Borel transform $B[{\cal{D}}^{(L)}_{PT}](z)$
can be found in Ref.\cite{r2}. It has the form
\begin{eqnarray}
&&B[{\MD}^{(L)}_{PT}](z)
=\sum_{n=1}^{\infty}\frac{A_{0}(n)-A_{1}(n)z_{n}}
{\Big{(}1+\frac{z}{z_{n}}\Big{)}^{2}}+\frac{A_{1}(n)z_{n}}
{\Big{(}1+\frac{z}{z_{n}}\Big{)}}\nonumber\\
&+&\sum_{n=1}^{\infty}\frac{B_{0}(n)+B_{1}(n)z_{n}}
{\Big{(}1-\frac{z}{z_{n}}\Big{)}^{2}}-\frac{B_{1}(n)z_{n}}
{\Big{(}1-\frac{z}{z_{n}}\Big{)}}
\end{eqnarray}
The expressions for the ${A}_{0,1}(n)$ and ${B}_{0,1}(n)$ are given in Ref.\cite{r2}.
For the Adler function in leading-$b$ approximation we see that there are single and double
poles in $B[{\MD}^{(L)}_{PT}](z)$ at positions $z=z_n$ 
and $z=-z_n$ with $z_n\equiv{2n}/b$ $n=1,2,3,\ldots$.
The singularities on the positive real semi-axis are the {\it infrared} renormalons, ${IR}_{n}$ and those on the
negative real semi-axis are {\it ultraviolet} renormalons, ${UV}_{n}$. We shall see that they correspond to
integration over the bubble-chain momentum $k^2$ in the regions $k^2<Q^2$ and $k^2>Q^2$, respectively.
The IR renormalon singularities lie on the Borel integration path along the positive real axis. This
leads to an ambiguous imaginary part which is structurally the same as a term in the OPE expansion in Eq.(8).
The ${IR}_{n}$ renormalon ambiguity is in one-to-one correspondence with non-logarithmic UV divergences present
in the ${(\Lambda^2/Q^2)}^{n}$ term of the OPE \cite{r7}.
OPE ambiguities and perturbative ambiguities can cancel once a definite regulation of the Borel integral
, for instance principal value (PV), has been chosen, and the PT and NP components are then separately
well-defined.
The (PV regulated) Borel integral may be evaluated in terms of exponential integral $Ei$ functions \cite{r2},
but notice that the Borel integral diverges for $Q^2<\Lambda^2$ , and potentially for $Q^2=\Lambda^2$ !
\begin{eqnarray}
&&{\MD}^{(L)}_{PT}(Q^2)=\nonumber \\
&&\sum_{n=1}^{\infty}[{z}_{n}{e}^{{z}_{n}/a(Q^2)}{\rm{Ei}}\left(\frac{z_n}{a(Q^2)}\right)
\nonumber \\
&\times&\left[\frac{z_n}{a(Q^2)}(A_0(n)-z_n{A}_{1}(n))-z_n{A}_{1}(n)\right]
\nonumber \\
&+&(A_0(n)-{z}_{n}A_1(n))]
\nonumber \\
&+&\sum_{n=1}^{\infty}{z}_{n}[{e}^{-z_n/a(Q^2)}{\rm{Ei}}\left(\frac{z_n}{a(Q^2)}\right)
\nonumber \\
&\times&\left[\frac{z_n}{a(Q^2)}(B_0(n)+{z}_{n}{B}_{1}(n))-{z}_{n}{B}_{1}(n)\right]
\nonumber \\
&-&({B}_{0}(n)+z_n{B}_{1}(n))]
\end{eqnarray}
This expression has the property that it is finite and continuous at $Q^2=\Lambda^2$
and freezes smoothly to a freezing limit of ${\MD}_{PT}^{(L)}(0)=0$. Similar behaviour is found for
GLS/polarized Bjorken and unpolarized Bjorken DIS sum rules, ${\MK}_{PT}^{(L)}(Q^2)$ and ${\MU}^{(L)}_{PT}(Q^2)$.
See \cite{r1} for their definitions. Interesting connections between these sum rules have been explored in Ref.\cite{r8}.
The finiteness and continuity are delicate. The $Ei$
functions potentially have a divergence proportional to $\ln{a(Q^2)}$ as $Q^2\rightarrow{\Lambda}^2$, but
the coefficient of this divergent term is
\begin{eqnarray}
-\sum_{n+1}^{\infty}{z}_{n}^{2}[A_1(n)+B_1(n)].
\end{eqnarray}
For ${\MK}^{(L)}_{PT}(Q^2)$ and ${\MU}^{(L)}_{PT}(Q^2)$ the equivalent coefficients are
$(-8+2+16-10=0)$ and $(8-6-2)=0$, respectively. There is a relation between IR and UV renormalon residues
which ensures the divergent term vanishes
\begin{equation}
{z}^{2}_{n+3}B_1(n+3)=-{z}_{n}^{2}A_1(n).
\end{equation}
This ensures that
\begin{equation}
\sum_{n=1}^{\infty}{z}_{n}^{2}[A_1(n)+B_1(n)]=0.
\end{equation}
Another similar relation is \cite{r2}
\begin{equation}
{A}_{0}(n)=-{B}_{0}(n+2).
\end{equation}
We shall show that these relations are underwritten by continuity of the characteristic
function in the skeleton expansion.\\

The one chain term in the QCD skeleton expansion can be written in the form \cite{r9} 
\begin{equation}
{\cal{D}}^{(L)}_{PT}(Q^2)=\int_{0}^{\infty}{dt}\;{\omD}(t)a({t}{Q}^{2})\;.
\end{equation}
Here ${\omD}(t)$ is the {\it characteristic function}. It satisfies the normalization condition
\begin{equation}
\int_{0}^{\infty}{dt}\;{\omD}(t)=1\;.
\end{equation}
$t\equiv k^2/Q^2$, and so one is integrating over the momentum flowing through the chain of bubbles.
${\omD}(t)$ and its first three derivatives are continuous \cite{r9} at $t=1$, and the integral divides into
an IR and UV part, corresponding to $k^2<Q^2$ and $k^2>Q^2$, respectively.
\begin{equation}
{\MD}^{(L)}_{PT}=\int_{0}^{1}{dt}\omD^{IR}(t)a(tQ^2)+\int_1^{\infty}{dt}\omD^{UV}a(tQ^2).
\end{equation}
The first term involving ${\omega}_{\cal{D}}^{IR}$ reproduces the IR renormalon contributions, and
the second term involving ${\omega}_{\cal{D}}^{UV}$ the UV renormalon contributions. For $Q^2>{\Lambda}^2$ the first integral
encounters the Landau pole in the coupling in the region of integration and requires regulation. If one uses a PV definition one can show by a change of variable
that the one-chain skeleton expansion result is exactly equivalent to the PV regulated Borel integral of Eq.(14),
and yields ${\MD}^{(L)}_{PT}(Q^2)$ as in Eq.(16). For $Q^2<{\Lambda}^2$ the first integral no longer requires regulation,
but the second UV term does. If one uses PV regulation one can show that the one-chain result is exactly equivalent
to a PV-regulated  modified Borel representation \cite{r1,r10},
\begin{eqnarray}
{\cal{D}}^{(L)}_{PT}({Q}^{2})=\int_{0}^{-\infty}{dz}\,{e}^{-z/a(Q^2)}B[{\cal{D}}^{(L)}_{PT}](z)\;.
\end{eqnarray}  
Note that the contour of integration now runs along the negative real axis in the Borel plane, and therefore
there is now an ambiguous imaginary part due to the UV renormalons. The ambiguity contributed by ${UV}_{n}$
is of the form ${(Q^2/\Lambda^2)}^{n}$. These ambiguities are associated with IR divergences of dimension-six
four-fermion operators associated with UV renormalons \cite{r11}, and this suggests that the NP OPE component for
$Q^2<\Lambda^2$ should be replaced by an expansion analogous to Eq.(8), but in powers of $Q^2/\Lambda^2$. 
Continuity of $\omD (t)$ and its first three derivatives at
$t=1$, and equivalently finiteness of ${\MD}^{(L)}_{PT}(Q^2)$ and its first three derivatives $d/d\ln Q$
at $Q^2=\Lambda^2$ is underwritten by the relation of Eq.(19), and by three additional
more complicated relations involving the ${A}_{0,1}$ and ${B}_{0,1}$ residues \cite{r1}. 
It is easy to show that the ambiguous imaginary part in ${\MD}^{(L)}_{PT}$ arising from IR renormalons
for $Q^2>\Lambda^2$, and UV renormalons for $Q^2<\Lambda^2$, can be written directly in terms of $\omD^{IR}$
and $\omD^{UV}$. For $Q^2>\Lambda^2$ one has
\begin{equation}
Im[{\MD}^{(L)}_{PT}(Q^2)]=\pm \frac{2\pi}{b}\frac{{\Lambda}^{2}}{Q^2}\omD^{IR}\left(\frac{\Lambda^{2}}{Q^2}\right),
\end{equation}
and for $Q^2<\Lambda^2$,
\begin{equation}
Im[{\MD}^{(L)}_{PT}(Q^2)]=\pm\frac{2\pi}{b}\frac{{\Lambda}^{2}}{Q^2}\omD^{UV}\left(\frac{\Lambda^{2}}{Q^2}\right).
\end{equation}
Continuity at $Q^2=\Lambda^2$ then follows from continuity of $\omD (t)$ at $t=1$. In principle the real part
of the OPE condensates are independent of the imaginary, but continuity and finiteness involve the set
of relations between ${A}_{0,1}$ and ${B}_{0,1}$ that we have just noted. Continuity naturally follows if we write 
${\MD}^{(L)}_{NP}(Q^2)$ in the form
\begin{equation}
\left({\kappa}\pm\frac{2\pi i}{b}\right)\int_{0}^{{\Lambda}^2/Q^2}{dt}\;\left({\omD}(t)+t\frac{d\omD(t)}{dt}\right).
\end{equation}
Here $\kappa$ is an overall real non-perturbative constant. If the PT component is PV regulated then one
averages over the $\pm$ possibilities for contour routing, combining with ${\MD}^{(L)}_{PT}$ one can then write down
the overall result for ${\MD}^{(L)}(Q^2)$.
The $Q^2$ evolution is fixed by the non-perturbative constant $\kappa$ and by $\Lambda$. Both the PT and NP components freeze smoothly to zero.\\

It is a pleasure to thank the organisers of ICHEP'06 for arranging such a stimulating
conference. Thanks also to Irinel Caprini, Jan Fischer, Andrei Kataev and Dimitri Shirkov
for some useful and entertaining discussions.

\end{document}